\documentclass[11pt,onecolumn,draftclsnofoot]{IEEEtran}
\usepackage{amsfonts}
\IEEEoverridecommandlockouts

\ifCLASSINFOpdf
\else
\fi
\usepackage{psfig}
\usepackage{epsf}
\usepackage{bm} 
\usepackage{slashbox}
\usepackage{booktabs}
\usepackage{pifont}
\usepackage{subfigmat} 
\usepackage{times,amsmath,color,amssymb,graphicx,geometry, epsfig,psfrag,subfigure,algorithm}
\usepackage{enumerate}
\usepackage{color}

\begin{document}
\title{ Mobile Edge Computing in Unmanned Aerial Vehicle Networks}
\author{\IEEEauthorblockN{Fuhui Zhou, \emph{Member, IEEE}, Rose Qingyang Hu, \emph{Senior Member, IEEE},\\
 Zan Li, \emph{Senior Member, IEEE},  Yuhao Wang, \emph{Senior Member, IEEE}}
\thanks{Fuhui Zhou is with the School of Information Engineering, Nanchang University, and with the State Key Laboratory of Integrated Services Networks, Xidian University, Xi¡¯an 710071, China. He is also with the Department of Electrical and Computer Engineering at Utah State University, Logan, UT 84322 USA (e-mail: zhoufuhui@ieee.org).

Rose Qingyang Hu is with the Department of Electrical and Computer Engineering at Utah State University, Logan, UT 84322 USA (e-mail: rose.hu@usu.edu).

Zan Li is with the Integrated Service Networks Lab of Xidian University, Xi'an, 710071, China (e-mail: zanli@xidian.edu.cn).

Yuhao Wang is with the School of Information Engineering, Nanchang University, P. R. China, 330031 (e-mail: wangyuhao@ncu.edu.cn).
}

\thanks{This work was supported in part by the National Natural Science Foundation
of China (61701214 and 61661028), in part by the Intel Corporation, and in part by the US National Science Foundation under Grant EARS1547312,  in part by the National Natural Science Foundation for
Distinguished Young Scholar under Grant 61825104, in part by the Key project of National Natural Science Foundation of China under Grant 61631015,  in part by the Open Foundation of The State Key Laboratory of Integrated Services Networks under Grant
ISN19-08, in part by the Excellent Youth Foundation of Jiangxi Province under Grant 2018ACB21012.}
}
\maketitle
\begin{abstract}
Unmanned aerial vehicle (UAV)-enabled communication networks are promising in the fifth and beyond wireless communication systems. In this paper, we shed light on three UAV-enabled mobile edge computing (MEC) architectures. Those architectures have been receiving ever increasing research attention for improving computation performance and decreasing execution latency by integrating UAV into MEC networks. We present a comprehensive survey for the state-of-the-art research  in this domain.  Important implementation issues are clarified. Moreover, in order to provide an enlightening guidance for future research directions, key challenges and open issues are discussed.
\end{abstract}
\IEEEpeerreviewmaketitle
\section{Introduction}
\IEEEPARstart{T}{he} emerging computation-intensive applications (e.g., automatic navigation, face recognition, unmanned driving, etc.) are  triggering a revolution of mobile applications and can significantly improve the quality of experience (QoE) of mobile users. However, they impose a great challenge on mobile devices, which normally have  a low computation capability and finite battery capacity. Mobile edge computing (MEC) has been identified as a promising technique to tackle this challenge and is receiving increasing research attention from both industry and academia  \cite{Y. Mao}. It enables mobile users to offload partial or complete computation-intensive tasks to MEC servers for computing.  Moreover, the computation performance of mobile users is significantly improved in a cost-effective and energy-saving manner since MEC servers are normally deployed close to  end users.

Recently, unmanned aerial vehicle (UAV)-enabled MEC architectures have been proposed by integrating UAV into MEC networks \cite{Y. Mao}, \cite{Y. Zeng1}. In those architectures, an UAV can be considered  as a user that has computation tasks to be executed, or a relay to assist users for offloading computation tasks,  or an MEC server for executing computation tasks.  Compared to the conventional terrestrial MEC networks, UAV-enabled MEC networks have several prominent advantages. They can be flexibly deployed in most scenarios even in wilderness, desert, and complex terrains, where the terrestrial MEC networks may not be conveniently and reliably established. Moreover, the computation performance can be improved since there is a  high possibility  that short-distance line-of-sight (LoS) links exist for computation tasks offloading and computation results downloading. Furthermore, the trajectory of the UAV can be optimized to further improve the user computation performance. They can be particularly helpful in the situations that conventional terrestrial MEC systems are destroyed by natural disasters.  There have been many leading companies (e.g.,  Google, Facebook, Amazon and Huawei) that have launched their projects to facilitate the application of UAV-enabled MEC networks \cite{Y. Mao}. It is envisioned that UAV-enabled MEC networks will be widely deployed when the cost of UAVs goes down sufficiently.

The research  and development on UAV-enabled MEC networks are still in their early stage and there have been very limited investigations along this direction \cite{N. Cheng}-\cite{J. Lyu}. In \cite{N. Cheng}, an UAV-enabled MEC architecture was presented, which the UAV can be flexibly deployed and scheduled to assist the communication, caching, and computing of the ground networks. The authors in \cite{N. Cheng} only presented the challenges of the proposed UAV-enabled MEC architecture. In order to realize the benefits  and  facilitate the wide applications of UAV-enabled MEC networks in practice, it is of crucial importance to understand their recent research advances as well as the technical  challenges.   Towards that end, this article presents a comprehensive survey on the state-of-the-art research efforts made in the domain of UAV-enabled MEC networks. The existing literatures  proposed three different UAV-enabled MEC network architectures based on the role that the UAV plays in the network. Moreover,  some key implementation issues of UAV-enabled MEC networks are highlighted. Furthermore, we shed light on the challenges in UAV-enabled MEC networks and discuss the open research issues.

The rest of the article is organized as follows. Section II presents the state-of-the-art studies on UAV-enabled MEC networks. Some key implementation issues are elaborated in Section III. Section IV discusses the challenges and open research issues. The paper is concluded in Section V.

\section{State of The Art}
In this section, the state-of-the-art studies on UAV-enabled MEC networks are presented.  The potential application scenarios and three UAV-enabled MEC architectures are first discussed.  The overview for the on-going research efforts in the domain of UAV-enabled MEC networks is provided.
\subsection{Application Scenarios}
\begin{figure}[!t]
\centering
\includegraphics[width=5.5in]{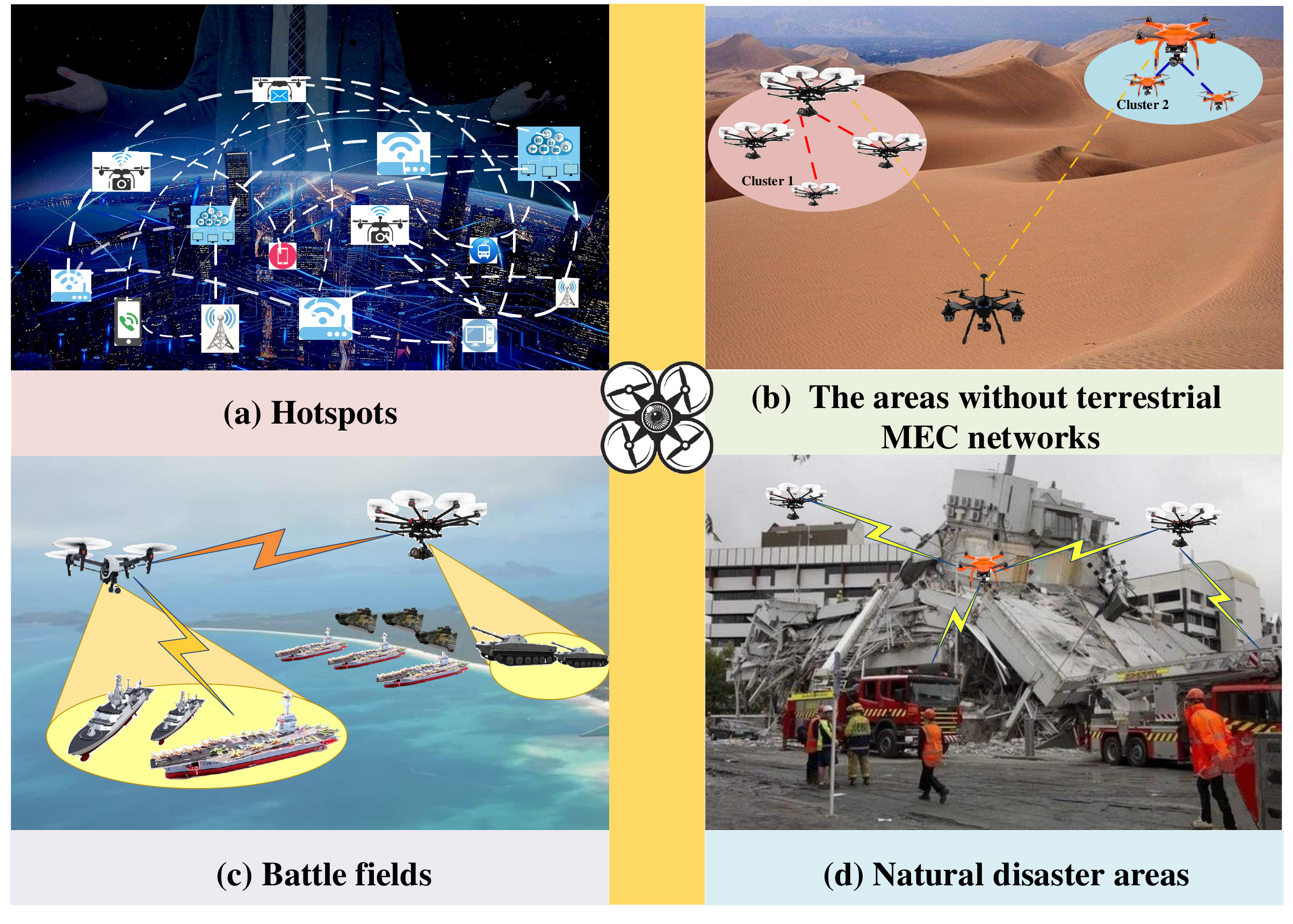}
\caption{Potential application scenarios of UAV-enabled MEC networks} \label{fig.1}
\end{figure}
\subsubsection{Hotspots} In hotspots, there can be massive users that demand computation-intensive services.  For example, in the city of Beijing, there can be more than one million peoples  simultaneously using the navigation applications during the rush hour. The UAV-enabled MEC networks can be exploited to assist the terrestrial MEC systems for executing the high volume computation tasks. The assistance from UAV can greatly decrease the outage probability and  improve user QoE.

\subsubsection{Outside the coverage of  terrestrial MEC networks} In many areas, such as wildernesses, deserts, and complex terrains, the government needs to keep monitoring the environment so that the corresponding strategies can be taken to protect the environment. In those areas, neither terrestrial MEC systems can be conveniently established nor the cost for establishing terrestrial MEC systems is  affordable. UAV-enabled MEC network becomes a desirable alternative.

\subsubsection{Battle fields} In battle fields,  there could be massive real-time computational tasks for special missions but it is extremely difficult to establish reliable terrestrial MEC systems.  For example, the locations and the  dynamics of the hostile forces need to be accurately estimated in a real time manner. UAV-enabled MEC networks can be  highly instrumental in these scenarios.

\subsubsection{Natural disasters} During natural disasters,  terrestrial MEC systems can be completely destroyed but many  rescue or reconstruction related  tasks  need to be computed and executed. In this case,  UAV-enabled MEC networks can play a very important role.
\subsection{Recent Advances}
\begin{figure}[!t]
\centering
\includegraphics[width=6in]{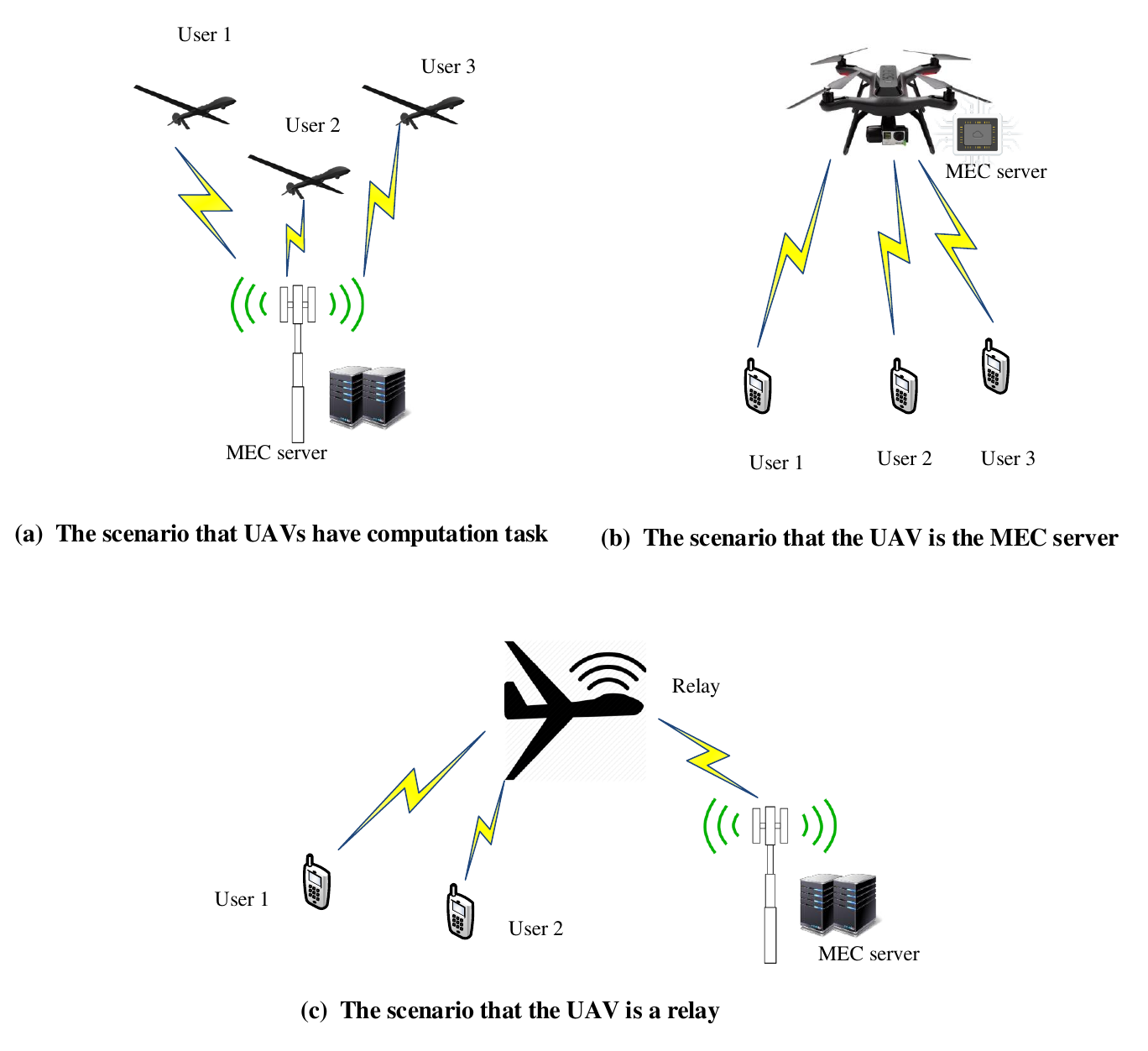}
\caption{Three UAV-enabled MEC architectures} \label{fig.1}
\end{figure}
Based on the role that the UAV plays, UAV-enabled MEC networks have three possible architectures.  The first architecture is presented in Fig. 2(a). The UAV functions as a user that needs to execute its own  computation tasks, such as trajectory  optimization.  UAV normally  has a finite battery capacity and  it may not be able to  perform extensive computational tasks. It will  need to offload its tasks to the ground MEC server for computing. For the second architecture shown in Fig. 2(b), the UAV functions as  the MEC server that helps the ground users to perform task computing after the ground users offload their computation tasks to the UAV. This  can  occur  in the areas without a terrestrial MEC network. In the third architecture shown in Fig. 2(c), the UAV works as a relay for assisting the users to efficiently offload their computation tasks to the MEC server.

Those three different UAV-enabled MEC architectures have their own application scenarios. The first  architecture is adaptable to the scenario where the UAV has limit computation capability but need to perform computation-intensive tasks. The second architecture is appropriate for  the scenario where the UAV has a considerable battery and computation capability and for the scenario where the areas do not have a terrestrial MEC network due to natural disaster.  The third  architecture is adaptable to the scenario where the UAV does not equip with a MEC server and  the offloading link quality between the user and the MEC server is poor. In this case, the UAV can work as a relay to help the user to offload its computation tasks to the MEC server. Fig. 3 shows the overview of the  recent research advances in UAV-enabled MEC networks. The details for the recent advances are presented in the following for all three architectures.

\begin{figure}[!t]
\centering
\includegraphics[width=5 in]{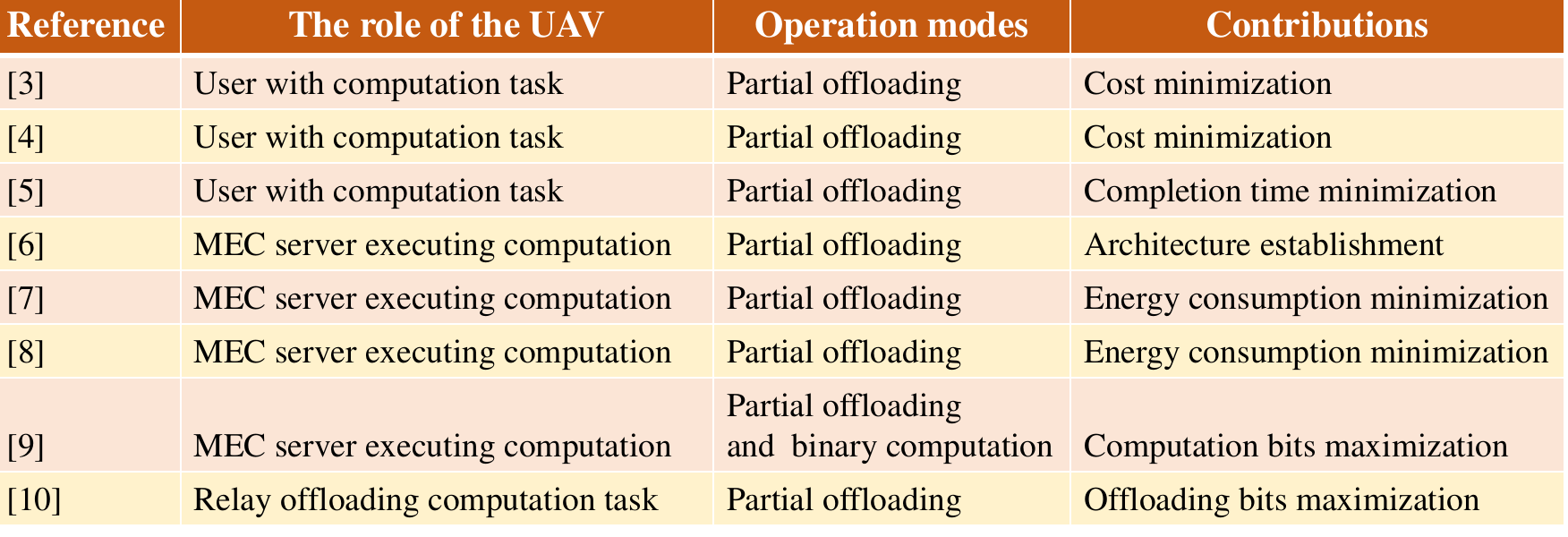}
\caption{Recent advances in the UAV-enabled MEC networks} \label{fig.1}
\end{figure}
In \cite{M. A. Messous}-\cite{X. Cao} the authors studied the first UAV-enabled MEC architecture. Specifically,  in \cite{M. A. Messous} and \cite{M. A. Messous1}, a theoretical methodology based on a sequential game was exploited to minimize the execution cost of the UAV-enabled MEC network. Three different types of players (drone, base station, and edge server) were considered in \cite{M. A. Messous} while  drone-only players with three different strategies were studied in \cite{M. A. Messous1}. The existence of the Nash Equilibrium was demonstrated in \cite{M. A. Messous}\cite{M. A. Messous1} and a good tradeoff between the energy consumption and the delay was achieved by using the proposed strategies. Note that UAVs in \cite{M. A. Messous} and \cite{M. A. Messous1} are assumed to hover in the sky and their trajectories are not optimized. Different from the works in \cite{M. A. Messous} and \cite{M. A. Messous1}, the authors in \cite{X. Cao} proposed a resource allocation strategy that jointly optimizes the  trajectory of the UAV  and the offloading time  in order to minimize the mission completion time of the UAV.

In \cite{N. Cheng}-\cite{F. Zhou}, the second UAV-enabled MEC architecture was studied, where the UAV is identified as the MEC server for providing computing services to the ground users. In particular, the authors in \cite{N. Cheng} established an UAV-enabled MEC architecture. The UAV can be flexibly deployed and scheduled to assist the communication, caching, and computing of the ground networks. In \cite{S. Jeong2}, the consumed energy of the UAV was minimized by jointly optimizing the offloading, downloading computing bits, and the local computing bits altogether while the total computation bits are guaranteed. Note that the energy consumed for the hover of the UAV was not considered and the trajectory of the UAV was not optimized. Based on the work in \cite{S. Jeong2},  the authors in \cite{S. Jeong} extended the energy minimization problem into a more practical UAV-enabled MEC network, where the energy of the UAV consumed for flying and the optimization of its trajectory were considered. Moreover, both orthogonal multiple access (OMA) and non-orthogonal multiple access (NOMA) schemes were studied. It was shown that the computation performance achieved with the NOMA scheme is better than that obtained by using the orthogonal frequency division multiple access (OFDMA) scheme. In \cite{F. Zhou}, the computation bits maximization problem was studied in UAV-enabled wireless powered MEC networks under both partial  and binary computation offloading modes. The central processing unit (CPU) frequencies, user offloading time, user offloading power and the trajectory of the UAV were jointly optimized. It was shown that the computation performance achieved under the partial offloading mode is superior to that obtained under the binary computation mode. Moreover, it was found that whether users choose to locally compute or offload their tasks to the UAV for computing depends on the tradeoff between the achievable computation bits and the operation cost. Furthermore, it was demonstrated that the optimization of the UAV trajectory  can significantly improve the computation performance.

For the third UAV-enabled MEC architecture, the authors in \cite{J. Lyu} proposed an optimal resource allocation scheme for maximizing the offloading bits under the max-min fairness criterion. The UAV was considered as a relay for offloading the computation bits of users to the MEC server. Moreover, a cyclical multiple access was defined in this paper. It was shown that  the computation performance can be significantly improved by optimizing  the UAV's trajectory and exploiting  the cyclical multiple access.

In order to extend  the coverage area, to provide diverse computation services with different scales of computation bits, and also to improve the operation efficiency, these three UAV-enabled MEC architectures can be deployed in practice under different scenarios. They can also co-exist and cooperate with each other in the same deployment area. The selection of the architecture depends on the scale of the computation bits and the function of the UAV. Specifically, when the UAV has its own computation tasks to be executed and if the UAV cannot complete all its computation tasks or if the UAV is not equipped with a computation circuit, the first architecture is a desirable choice. When the UAV has a strong computation capacity and does not have any its own computation tasks, the second architecture can be selected for helping the ground users to complete their computation tasks. Finally, the third architecture is a desirable candidate for cooperating with users to efficiently offload their computation tasks when the UAV only  has communication circuits and cannot compute. So far the research on these three architectures is still very limited.

\section{Technical Details for implementation}
\begin{figure}[!t]
\centering
\includegraphics[width=6 in]{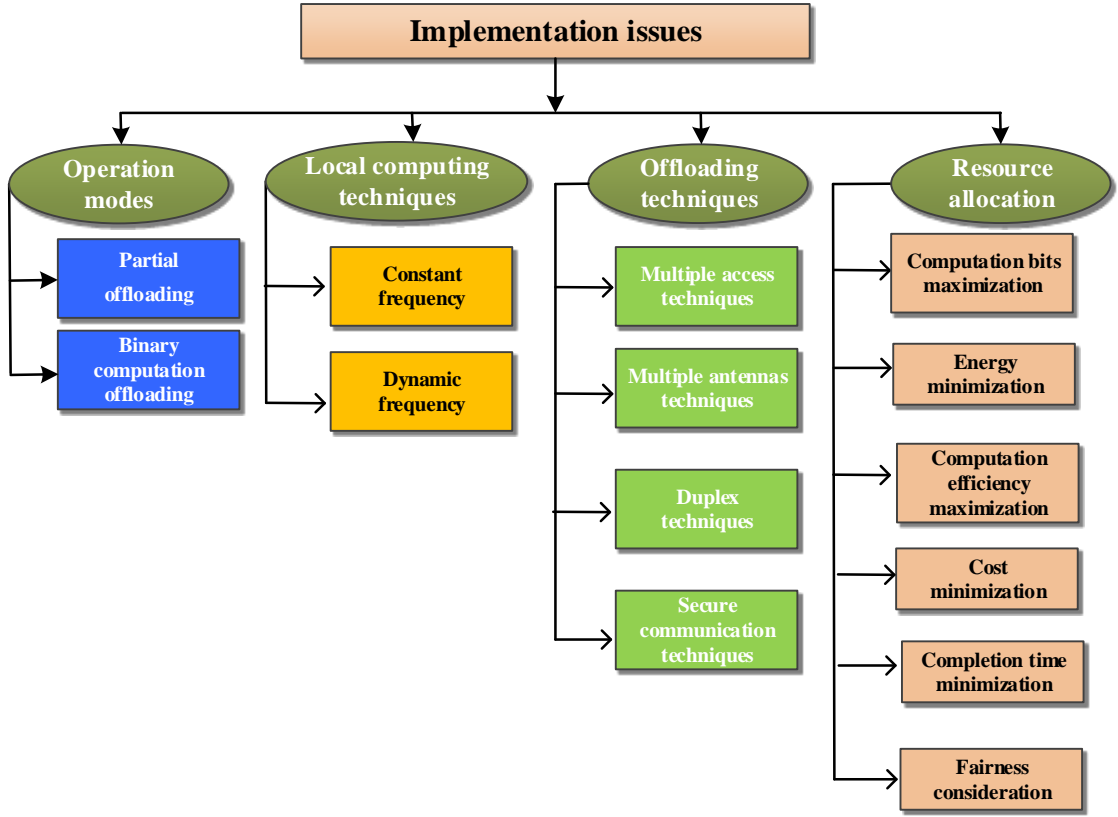}
\caption{Implementation Details} \label{fig.1}
\end{figure}
Some key implementation elements in UAV-enabled MEC networks are highlighted in Fig. 4.
\subsection{Operation Modes}
In UAV-enabled MEC networks, there can be  two operation modes, namely, partial offloading mode and binary computation mode. For the  partial offloading mode,  the computation task is partitioned into two parts. One part is offloaded to the MEC server for computing while the other part is locally computed \cite{M. A. Messous1}, \cite{S. Jeong2}-\cite{F. Zhou}. For example, when the face recognition task is performed,  low computation tasks such as to decide the existence of a human being  can be locally computed.  High computation tasks  such as characteristic recognition and facial feature recognition, can be offloaded to MEC for computing. For the binary computation mode,  each computation task has to be executed as a whole \cite{F. Zhou}.  It can be either executed locally or completely offloaded. For example, when the estimation of the channel state information (CSI) is performed,  the raw correlated data samples need to be computed altogether for improving the estimation accuracy.

These two operation modes have their respective  pros and cons. Specifically, under the partial offloading mode, the UAV can dynamically and flexibly allocate its communication resources for offloading computation tasks and its computation resources for performing local computation based on the CSI. Under the binary computation mode, the flexibility of the resource allocation scheme can be limited since the UAV cannot simultaneously perform task offloading and local computing. Thus, the computation performance achieved under the  partial offloading mode could be  better than that obtained under the binary computation mode \cite{F. Zhou}.  However, a more complex circuit and protocol is required under the partial offloading mode. Thus, the selection of the operation modes depends on the structure of the UAV and the characteristic of the computation task.
\subsection{UAV Computing Techniques}
There can be two local computing techniques in  UAV-enabled MEC networks based on the UAV CPU frequency. When the UAV computation circuit has a fixed CPU frequency, the local computation is performed with a constant rate \cite{M. A. Messous1}. If the UAV adopts a dynamic voltage and frequency scaling technique, it can adjust the CPU frequency  based on the scale of the computation tasks \cite{S. Jeong}, \cite{F. Zhou}.  The dynamic local computing technique can achieve a better performance than the constant local computing technique in terms of the energy consumption,  throughput,  and latency at the cost of a higher complexity of the computing circuit.
\subsection{Offloading Techniques}
For computation offloading, the communication techniques used for the terrestrial MEC networks can also be applied in the UAV-enabled MEC networks.  Further we can exploit more UAV provided functions for improving the conventional offloading techniques.
\subsubsection{Multiple access techniques}
Multiple access techniques can be classified into two categories, orthogonal multiple access (OMA) and non-orthogonal multiple access (NOMA). The typical OMA techniques in the UAV-enabled MEC networks are OFDMA and time division multiple access (TDMA), which have been extensively studied in the past \cite{X. Cao}\cite{S. Jeong2}.  In \cite{J. Lyu}, the authors proposed an OAM based cyclical multiple access (CMA) technique that schedules users to communicate with the UAV in a cyclical time-division manner when the UAV is close to the user.  It exploits the channel condition to improve the offloading efficiency. It was shown  that the offloading throughput achieved with CMA is larger than that obtained with TDMA.   In NOMA, multiple UAVs or users can communicate with the MEC server using the same physical resource.  NOMA can improve the user connectivity and spectral efficiency compared with OMA.  In \cite{S. Jeong2}, NOMA was applied to serve multiple users to offload their computation tasks in the uplink and for the UAV to send its computation results in the downlink. It was shown that NOMA can provide a higher computation performance gain than  OMA.
\subsubsection{Multiple antennas techniques} Multiple antenna techniques can improve the offloading efficiency by exploiting  the space diversity gain. UAV can be equipped  with multiple antennas to communicate with users or the MEC and vice versa \cite{J. Zhao}.   Different from the conventional beamforming design in the terrestrial MEC networks, the effect of the trajectory on the beamforming design should be considered. It was shown in \cite{J. Zhao} that UAV flying direction  has a big impact on the beamforming design and on the achievable performance.
\subsubsection{Duplex techniques} The  task offloading process and the outcome downloading process cannot be performed at the same time with half duplex technologies  while these two processes can be simultaneously conducted when  full duplex is applied.  A user can offload its computation task to the UAV while the UAV can transmit the computation results to the user at the same time with full duplex. Evidently  the offloading efficiency of the full duplex technique is higher than that of the half duplex technique at the cost of a more complex protocol and circuit design \cite{S. Mao}. Moreover, the interference between the offloading process and the downloading process should be effectively controlled in full duplex.

\subsubsection{Security}  Although it is recognized  that MEC network can provide better security  than the cloud computing network owing to the short distance between users and the MEC server, secure techniques are still needed in the UAV-enabled MEC networks. There can be  a high possibility that  LOS communication links exist between the UAV and eavesdroppers (if there are any) so that   eavesdroppers can intercept the confidential computation tasks and the results.   Even worse,  since traditional cryptographic techniques through data-encryption need to add the computation overhead for achieving security, they are inappropriate  in UAV-enabled MEC networks  due to latency  as well as power concerns.  As alternatives,  physical-layer security techniques  in UAV-enabled MEC networks gain attention since they leverage the physical layer characteristics of the wireless channels to realize secure communications \cite{J. Xu}. When the CSI between UAV and the legitimate user  is better than that between the UAV and eavesdropper, the offloading is secure. Moreover, in  UAV-enabled MEC networks, the UAV trajectory can be optimized so that the UAV can stay close to the legitimate user and keep itself far away from the eavesdropper.

\subsection{Resource Allocation}
Resource allocation is of vital importance in  UAV-enabled MEC networks due to UAV battery concern and trajectory constraint.  A good resource allocation scheme can not only improve the computation performance, but also realize an efficient and economical operation of the UAV-enabled MEC systems. Moreover, an appropriate trajectory can be designed to achieve a good tradeoff between the computation performance and the operation cost of the UAV.  Most of the existing works have  focused on designing  resource allocation strategies  for realizing different objectives by using diverse optimization techniques \cite{M. A. Messous}-\cite{J. Lyu}. Different from the resource allocation in conventional terrestrial MEC networks, resource allocation in UAV-enabled MEC networks should consider resources involved in three processes, namely, computation task offloading, local computing, and UAV's flying.  The resources to be optimized  in the computation task offloading are the communication resources, such as communication bandwidth, offloading power, and offloading time, etc. In the local computing,  the resources include the CPU frequency and the computation time. For the UAV's flying process, the resources to be optimized are the trajectory, the fly speed and the accelerated velocity of the UAV.  Resource allocation in the UAV-enabled MEC networks can be designed to achieve different objectives, such as computation bits maximization, energy minimization, computation efficiency maximization, cost minimization, completion time minimization, and  fairness consideration. These objectives are respectively  presented as follows.

\subsubsection{Computation bits maximization} This objective  aims to maximize the number of the total computation bits by offloading and local computing \cite{F. Zhou}. It can directly reflect the computation performance of UAV-enabled MEC networks. When UAV is the user or the MEC server,  those three processes  should be jointly optimized to maximize the total computation bits while  when UAV functions as a relay,  offloading process and the UAV's fly process are the focus.

\subsubsection{Energy minimization} The energy consumed in UAV-enabled MEC networks comes from the above-mentioned three processes. In the local computing process, the consumed energy depends on the CPU frequency and the computation time \cite{S. Jeong}. In the offloading process, the consumed energy includes transmit power and the offloading time. And the energy consumed in the fly process should consider the UAV's weight, speed, accelerated velocity and fly time. Two different models for the consumed energy in the fly process have been proposed in \cite{S. Jeong2} and \cite{Y. Zeng}, respectively. A resource allocation scheme was designed to minimize the total energy consumed during these three processes in \cite{S. Jeong2}.
\subsubsection{Computation efficiency maximization} The computation bits maximization or the energy minimization overemphasizes the importance of one metric and cannot achieve a good tradeoff between the computation bits and the energy consumption. In contrast, the computation efficiency maximization aims to maximize the computation bits per Joule energy and can realize a good tradeoff between those two metrics \cite{H. Shi}.  Note that computation efficiency characterizes the efficiency during the computation process while energy efficiency characterizes the efficiency during the information transmission process. They are two different definitions.
\subsubsection{Cost minimization} In \cite{M. A. Messous} and \cite{M. A. Messous1}, the authors defined the cost of UAV-enabled MEC networks by considering the energy overhead and the execution latency. A good tradeoff between the energy overhead and the execution latency can be achieved  under cost minimization objective \cite{M. A. Messous} and \cite{M. A. Messous1}.

\subsubsection{Completion time minimization} The completion time is defined as the bigger value among the two, the local computation time and the offloading time,  under the partial offloading mode,  or as the total time of the local computation time and the offloading time under the binary computation mode. This optimization objective is to minimize the mission completion time while the minimum computation bits requirement is satisfied \cite{X. Cao}.
\subsubsection{Fairness consideration} Fairness should be considered when multiple users compete the resources \cite{S. Mao}.   Five fairness criteria are normally used, namely, max-min fairness,  proportional fairness,  harmonic fairness, $\alpha$ fairness, and $\beta$ fairness \cite{H. Shi}.

\section{Challenges and Open Issues}

 UAV-enabled MEC networks  are considered promising technologies for their great potentials in enhancing  connectivity and  improving user QoS.   We are still facing critical  challenges  in order to facilitate their wide range of applications.

\textbf{Resource allocation in multiple-user multiple-UAV scenarios:} Although many studies  have been conducted in designing resource allocation schemes \cite{M. A. Messous}-\cite{J. Lyu}, those schemes were mainly  proposed for networks with either one  UAV or one user. Since the operation time and battery of the UAV are limited and normally a large number of  users need to be served in the geographical coverage area, it is important yet challenging  to design efficient resource allocation schemes for UAV-enabled MEC networks with  multiple users and  multiple UAVs. The problem of grouping users with UAVs is typically an integer non-convex optimization problem. Moreover, the trajectories of multiple UAVs should be jointly optimized in order to expand the converge area and improve the computation performance, making the problem even more difficult.

\textbf{Operation selection under the binary computation mode:} In the binary computation mode,  binary selection variables related to either local computation or offloading tasks make the   operation selection problem a challenging mixed integer non-convex optimization problem.  The problem becomes even more difficult when there exist multiple users.

\textbf{Beamforming} As mentioned in Section III, the effect of the flying  direction on the beamforming design should be considered. This is challenging since the direction of the UAV changes dynamically over time.  Issues related to  the dynamic  beamforming design for the UAV-enabled MEC networks call for extensive  research.

\textbf{Security issues:}  To the best of our knowledge, there have been no studies on the  security issues in UAV-enabled MEC networks. As mentioned in Section III,  physical-layer security techniques provide a promising direction or pursue  in UAV-enabled MEC networks.

\textbf{Computation efficiency:} Since UAV has a limited energy support from battery, computation efficiency is of vital importance in UAV-enabled MEC networks. However, it is extremely challenging to tackle the computation efficiency problems due to the intractable fractional forms, especially  when the binary computation mode is considered.

\textbf{Cooperation between UAV-enabled and terrestrial MEC networks:}  In some scenarios like hotpot, computation tasks  need to be executed for meeting the computation service requirements of a large number of  users.  It is interesting to investigate cooperative techniques between UAV-enabled and terrestrial MEC networks. Two issues  need to be tackled for realizing this cooperation. How to cooperatively  allocate the communication resources for both UAV-enabled and terrestrial MEC networks should be studied.  The interference between  UAV-enabled and terrestrial MEC networks needs to be coordinated when they operate on the same frequency band.

\textbf{The application of machine learning:} In order to achieve intelligent control of UAVs and enhance the computation performance of UAV-enabled MEC networks, the application of machine learning is promising. However, how to exploit multiple-agent reinforcement learning for scheduling multiple UAV is challenging, especially when there are no cooperation among UAVs.

\begin{figure}[!t]
\centering
\includegraphics[width=6 in]{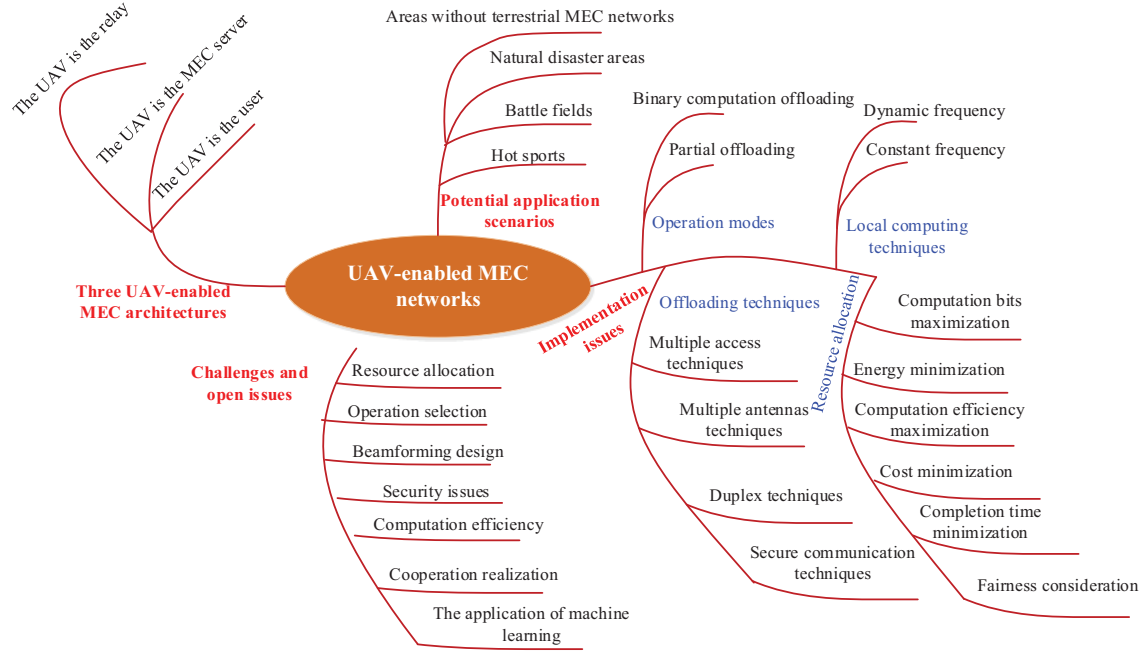}
\caption{Comprehensive summary on discussions} \label{fig.1}
\end{figure}
Fig. 5 gives a comprehensive summary on the potential application scenarios, implementation issues, challenges, and open issues of UAV-enabled MEC networks.
\section{Conclusions}
UAV-enabled MEC networks are promising for improving the computation performance and reducing the execution latency. In order to realize the benefits and the challenges of UAV-enabled MEC networks, this article presented a comprehensive survey on the recent advances made in this domain.  Detailed implementation issues  were highlighted for facilitating the wide range of  applications of UAV-enabled MEC networks. Key challenges and open issues were discussed in order to provide an enlightening guidance for future research directions.  The research on the UAV-enabled MEC networks is in its early stage. Extensive research efforts are needed to bring the technology to its maturity.

\section{Biographies}
Fuhui Zhou (M'16) received the Ph. D. degree from Xidian University, Xi¡¯an, China, in 2016. He is an associate Professor with School of Information Engineering, Nanchang University. He is now a Research Fellow at Utah State University. His research interests focus on cognitive radio, green communications, edge computing, machine learning, NOMA, physical layer security, and resource allocation. He serves as an Associate Editor of IEEE Systems Journal, IEEE Access and Physical Communications. He also serves as co-chair of IEEE ICC 2019 and IEEE Globecom 2019 workshop on \lq\lq Advanced Mobile Edge /Fog Computing for 5G Mobile Networks and Beyond\rq\rq.  (e-mail: zhoufuhui@ieee.org)\\

Rose Qingyang Hu (S'95-M'98-SM'06)  is a Professor in the Electrical and Computer Engineering Department and Associate Dean for research of College of Engineering  at Utah State University. She also directs Communications Network Innovation Lab at Utah State University. Her current research interests include next-generation wireless system design and optimization, Internet of Things, Cyber Physical system, Mobile Edge Computing, V2X communications, artificial intelligence in wireless networks, wireless system modeling and performance analysis. She is an IEEE Communications Society Distinguished Lecturer Class 2015-2018.  She was a recipient of prestigious Best Paper Awards from the IEEE GLOBECOM 2012, the IEEE ICC 2015, the IEEE VTC Spring 2016, and the IEEE ICC 2016.  Prof. Hu is senior member of IEEE and a member of Phi Kappa Phi Honor Society. (e-mail: rose.hu@usu.edu)\\

Zan Li (S'14) received the B.S. degree in communications engineering, and the M.S. and Ph.D. degrees
in communication and information systems from Xidian University, Xi¡¯an, China, in 1998, 2001, and
in 2004, respectively. She is currently a Professor with the State Key Laboratory of Integrated Services
Networks, Xidian University. Her research interests focus on topics in wireless communications and signal processing, including weak signal detection, spectrum sensing, co-operative communication. (e-mail: zanli@xidian.edu.cn)
\\

Yuhao Wang (S'14) is currently a Professor with the Cognition Sensor Network Laboratory, School of Information Engineering, Nanchang University (NCU), Nanchang, China. He is the Deans of the Information Engineering School and the Artificial Intelligence Industry Institute, NCU, and also is the Head of Jiangxi Embedded Systems Engineering Research Center. Since 2016, he is an IET Fellow. His current research interests include wideband wireless communication and radar sensing fusion system, channel measurement and modeling, nonlinear signal processing, smart sensor, image and video processing, and machine learning. (e-mail: wangyuhao@ncu.edu.cn) \\
\end{document}